# The silicon vacancy centers in SiC: determination of intrinsic spin dynamics for integrated quantum photonics

Di Liu[1], Florian Kaiser[2,#], Vladislav Bushmakin[1], Erik Hesselmeier[1], Timo Steidl[1],

Takeshi Ohshima[3,4], Nguyen Tien Son[5], Jawad Ul-Hassan[5], Öney O. Soykal[6,*]

and Jörg Wrachtrup[1]

**Affiliations:**

[1] 3rd Institute of Physics, IQST, and Research Center ScoPE, University of Stuttgart, Germany

[2] Luxembourg Institute of Science and Technology, Luxembourg

[3] National Institutes for Quantum Science and Technology (QST), 1233 Watanuki, Takasaki, Gunma 370-1292, Japan

[4] Department of Materials Science, Tohoku University, 6-6-02 Aramaki-Aza, Aoba-ku, Sendai 980-8579, Japan

[5] Department of Physics, Chemistry and Biology, Linköping University, Sweden

[6] Booz Allen Hamilton Inc., USA (current affiliation: Photonic Inc., CA)

**Email correspondence:**

* Öney O. Soykal, osoykal@photonic.com
# Florian Kaiser, florian.kaiser@list.lu

**Abstract**

The negatively-charged silicon vacancy center ($V_{Si}^-$) in silicon carbide (SiC) is an emerging color center for quantum technology covering quantum sensing, communication, and computing. Yet, limited information currently available on the internal spin-optical dynamics of these color centers prevents us achieving the optimal operation conditions and reaching the maximum performance especially when integrated within quantum photonics. Here, we establish all the relevant intrinsic spin dynamics of negatively charged $V_{Si}^-$ center in 4H-SiC by an in-depth electronic fine structure modeling including intersystem-crossing and deshelving mechanisms. With carefully designed spin-dependent measurements, we obtain all previously unknown spin-selective radiative and non-radiative decay rates. To showcase the relevance of our work for integrated quantum photonics, we use the obtained rates to propose a realistic implementation of time-bin entangled multi-photon GHZ and cluster state generation. We find that up to 3-photon GHZ/cluster states are readily within reach using the existing nanophotonic cavity technology.

**Introduction**

Solid-state spin qubits based on color centers in wide bandgap semiconductors are one of the leading platforms for quantum networks, information processing, and sensing[1–4] owing to their robust spin-optical properties and long coherence times. Among many available host materials[5–7], silicon carbide (SiC) particularly stands out as a wafer-scalable material with well-established isotopic engineering and compatibility with today's complementary metal-oxide-semiconductor (CMOS) microfabrication technology providing a path towards scalable systems. Rapid progress with spin qubits in SiC has already been made including the milestone demonstrations of millisecond spin coherence times at room temperature[8,9], high-fidelity spin and optical control[10], coherent spin-photon interfaces[11], entanglement with nuclear spin registers[12], and single-shot charge readout [13]. Capitalizing on the SiC's material advantages, steps towards scalability have also been implemented through successful integration of negatively charged silicon vacancy color centers ($V_{Si}^-$) into nanophotonic waveguides[14] and resonators[15–17], the latter one being compatible with SiC-on-insulator processing[18]. These proof-of-concept demonstrations identified the cubic lattice site $V_{Si}^-$ in 4H-SiC as a strong contender for quantum applications based on a dense integration of multiple color centers on one chip. Further progress towards fully scalable integrated solutions with $V_{Si}^-$ necessitates a complete understanding of its intrinsic spin-optical dynamics to guide the engineering efforts of cavity-emitter coupling and optimization of spin and optical properties. It will also provide the critical insights and metrics necessary for developing realistic quantum network applications.

In this paper, we reveal the comprehensive internal spin dynamics of the cubic lattice site $V_{Si}^-$ in 4H-SiC (V2 center) by theoretical characterization of its electronic structure. The theory is confirmed by our experimental investigations, which include the measurements of spin-selective excited state (ES) lifetimes, ground state spin initialization via resonant and off-resonant laser excitation, as well as probing the intricate dynamics within the metastable state (MS) manifolds via spin manipulation combined with a novel delayed pulse measurement. In this way, we determine all the spin-dependent radiative and non-radiative transition rates and identify the intersystem crossing (ISC) mechanism which all play a crucial role in defining realistic protocols for scalable quantum network applications. To showcase this, we develop a protocol for generating time-bin entangled multi-photon Greenberger-Horne-Zeilinger[19] (GHZ) and cluster states[20], which are particularly important for quantum network applications, one-way quantum computation, and quantum repeaters. Using the involved intrinsic transition rates and ISC mechanism, we provide estimates for quantum efficiencies, state fidelities, optimal pulse timings, and minimum requirements for a cavity enhancement of radiative lifetimes. The approaches and insights developed here are also directly applicable to other color centers and their benchmarking for specific applications.

**Electronic fine structure of $V_{Si}^-$ in 4H-SiC**

The crystal structure of 4H-SiC allows for two nonequivalent lattice sites, so-called hexagonal (h) and cubic (k) sites, to be occupied by the deep center $V_{Si}^-$ defect. Defects belonging to the h- and k-sites, referred to as the V1 and V2 centers, have distinct optical resonant excitation signatures at zero-phonon line (ZPL) wavelengths of 862 and 916 nm, respectively. Five active electrons present in V1 and V2, originating from the four $sp^3$ dangling bonds surrounding the vacancy and an additional captured negative charge, result with optically active ground and excited states in a Kramer's degenerate $S$=3/2 spin quartet configuration.

Both V1 and V2 have a local symmetry that belongs to the $C_{3v}$ double point group that is only slightly distorted from the cubic $T_d$ symmetry. In the case of V2, this distortion is stronger because of the additional next nearest neighbor silicon atom present along the c-axis for k-sites. As we will show in this work, this leads to a completely different spin-optical dynamics of the V2 center compared to the formerly studied V1 center[21].

To investigate the dynamics of the V2 center (from now on dubbed as $V_{Si}^{-}$ center), we build a theoretical framework upon symmetry adapted molecular orbitals (MOs) and localized many body $sp^3$ electronic wavefunctions[22]. Combining the theory and our experimental measurements, we reveal all the spin-dependent radiative and non-radiative transition rates, and as a result the intrinsic optical and spin dynamics of $V_{Si}^{-}$. The comprehensive $V_{Si}^{-}$ electronic fine structure is shown in Fig. 1a, in terms of MOs obtained by the Density Functional Theory (DFT) after group theoretical analysis and considerations of many-body spin-orbit, spin-spin, and exchange interactions.

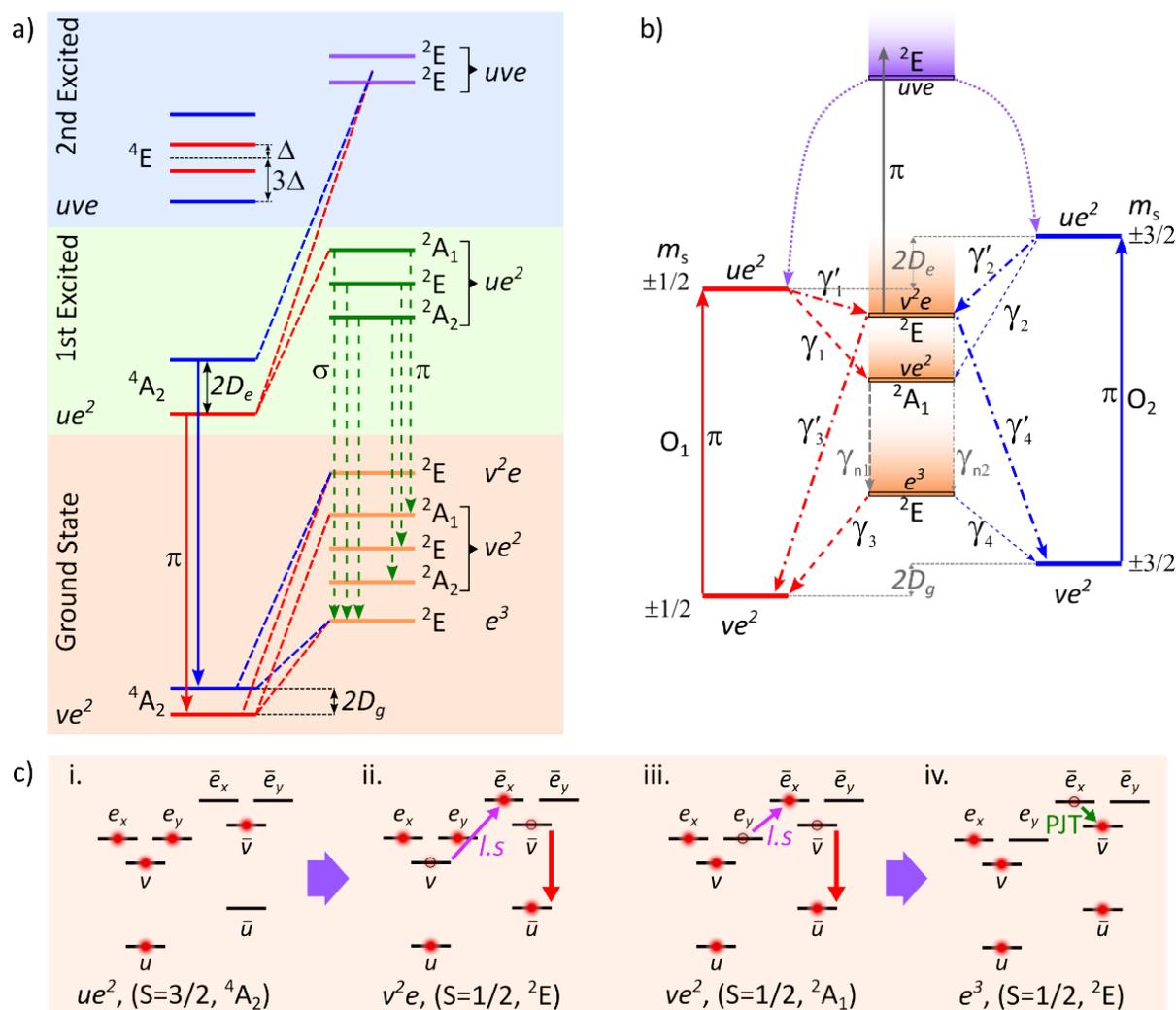

**FIG. 1. Energy levels and the intersystem crossing mechanism of the $V_{Si}^-$ center.**

**a)** Electronic fine structure of $V_{Si}^-$ center including the spin-orbit direct coupling between the quartets and the doublets.

**b)** The spin-selective ISC channels enabled by the spin-orbit and vibronic coupling.

**c)** Molecular orbital configurations of the states involved in the ISC. Bar (no bar) over the orbitals represents the spin majority (minority) channel within the spin polarized MO (DFT) picture. Starting from the (i) configuration, the configurations (ii) and (iii) can be reached by spin-orbit (purple arrows) enabled transitions accompanied by a fast optical or phonon assisted relaxation from the $\bar{v}$ orbital to the $\bar{u}$ orbital (red arrow). Configuration (iii) can also transition into (iv) enabled by the pseudo-Jahn Teller effect (green solid arrow).

The ground state (GS) manifold contains a spin-quartet state in an orbital-singlet ($^4A_2$ : $ve^2$) configuration and five metastable spin-doublet states. These doublets consist of two orbital-singlet ($^2A_1$, $^2A_2$ : $ve^2$) and three orbital-doublet ($^2E$ : $ve^2$, $e^3$, $v^2e$) configurations. The first excited state manifold differs from the ground state solely in the $u$ orbital, replacing $v$ ($ve^2 \rightarrow ue^2$). This leads to a spin-quartet orbital-singlet state ($^4A_2$ : $ue^2$) and three metastable spin-doublet states. These doublets consist of two orbital-singlets ($^2A_1$ : $ue^2$, $^2A_2$ : $ue^2$) and an orbital-doublet ($^2E$ : $ue^2$) configurations. The charge distribution localized on the nearest neighbor basal plane carbon atoms has an opposite parity for the $u$ and $v$ MOs, making the transition dipole moment ($\mu_{ve^2 \rightarrow ue^2}$) less sensitive to fluctuations in non-axial local electric field and strain, even though $V_{Si}^-$ defect lacks an inversion symmetry. This explains the experimentally observed high spectral stability of $V_{Si}^-$ center during a continuous resonant optical excitation[14,16]. The second excited state manifold includes another optically active spin-quartet orbital-doublet ($^4E$ : $uve$), which is split by the spin-orbit coupling, as well as two metastable spin-doublet orbital-doublet states ($^2E$ : $uve$). The symmetry-allowed spin-orbit coupling channels are indicated in Fig. 1a (dashed lines and arrows).

Based on the described fine-structure and the allowed coupling channels, we develop a simplified and a fully-equivalent energy level model of the $V_{Si}^-$ center as shown in Fig. 1b. It consists of Kramer's degenerate ground and excited states for the spin subspaces $m_s = \pm 1/2$ and $m_s = \pm 3/2$, and three effective metastable states -which can be further reduced to two as we show in the analysis below. The spin-conserving optical transitions between ground and excited states are denoted by $O_1$ and $O_2$ for each spin subspace whereas the additional radiative and intersystem-crossing (ISC) rates are denoted by $\gamma_i$. In this work, we use resonant and off-resonant optical excitation methods to extract all the transition rates shown in Fig. 1b, and we will show that the results are in excellent agreement with our theoretical calculations. The intrinsic spin dynamics of the $V_{Si}^-$ defect under optical illumination is governed by the radiative transitions between states of the same spin multiplicity as well as nonradiative ISCs between states of different spin multiplicity. In the case of optical excitation of the first excited state, the ISC is established by two processes involving:

i.) Transitions from the optical ($^4A_2 : ue^2$) first excited spin-quartet state to energetically higher metastable spin-doublet states ($^2E : v^2e$; $^2A_1 : ve^2$).

ii.) Transitions from the metastable spin-doublet states ($^2E : e^3, v^2e$) to the ground spin-quartet ($^4A_2 : ve^2$) state.

Both processes are mediated by a combination of spin-orbit (spin-lowering/raising) and electron-phonon (spin-conserving) interactions. The ISC mechanisms involved in these transitions can be explained intuitively using the simpler MO picture of the Density Functional Theory (DFT) as shown in Fig. 1c.

**ISC mechanism**

The upper ISC mechanism is governed by the $\gamma_{1,2}$ and $\gamma'_{1,2}$ rates (ES to MSs) in Fig. 1b. It consists of transitions from the $^4A_2 : ue^2$ spin quartet state to the $^2E : v^2e$ and $^2A_1 : ve^2$ spin-doublets, being assisted by both the spin-orbit coupling and the electron-phonon interaction. Consider the system to be initially in the excited state $ue^2$ spin-quartet configuration, as shown in the inset (i) of Fig. 1c.

From here, and as shown in the inset (ii) of Fig. 1c, the spin-orbit coupling ($\alpha_y l_y s_y$) can promote an electron from the $v$ orbital to the $\bar{e}_x$ orbital. This is followed by a fast spin-conserving decay of a second electron from the $\bar{v}$ orbital to the $\bar{u}$ orbital via either a photon emission or a phonon relaxation process. This then leads to the $v^2e$ spin-doublet configuration.

The inset (iii) of Fig. 1c shows an alternative pathway from the configuration in (i). Here, the spin-orbit coupling ($\alpha_z l_z s_z$) can promote an electron from the $e_y$ to the $\bar{e}_x$ orbital. Again, this is followed by the $\bar{v} \to \bar{u}$ decay, resulting in the $ve^2$ spin-doublet. In fact, due to the stronger spin-orbit coupling along the c-axis ($\alpha_z > \alpha_{x,y}$), the latter process (iii) is expected to be faster than (ii), which will be confirmed by our experimental investigations.

As shown in the inset (iv) in Fig. 1c, we must also consider that the $ve^2$ and the $e^3$ spin-doublet states can be vibronically coupled together by the $e$-symmetry acoustic phonons ($^2A_1 \times e \times {}^2E$) via the pseudo-Jahn Teller (PJT) effect. This allows the $ve^2$ spin-doublet of (iii) to transition into the $e^3$ spin-doublet state at a relatively fast dynamic relaxation rate ($\gamma_{n1} \gg \gamma_{1,2}$) which redistributes most of the population into the $e^3$ spin-doublet during an optical excitation cycle.

**Deshelving mechanism**

Subsequently, the non-radiative transition of the $e^3$ spin-doublet state to the ground $ve^2$ spin-quartet state forms the basis of the lower ISC mechanism represented by the effective $\gamma_{3,4}$ rates (MS to GS). In the MS manifold, the $ve^2$ doublets experience ultra-fast relaxation to the $e^3$ doublets ($\gamma_{n1}$), which is why we can combine them into a single effective spin-doublet state. On the other hand, the $v^2e$ state lacks a similar fast relaxation path ($\gamma_{n2} \sim 0$), thus resulting in low non-radiative decay rates $\gamma'_{3,4}$ to the ground state. Interestingly, the rather long lifetime of the $v^2e$ state can result in a sizeable optical excitation rate into the higher-lying $uve$ spin-doublet state (Fig. 1b). In this previously unknown deshelving process, an electron is promoted from the $\bar{u}$ to the $\bar{v}$ orbital. Here, the optical excitation is required to have the same polarization ($\pi$) and similar energy as the $O_1$ and $O_2$ transitions (only differing by the electron exchange correlations). The cycle is then completed by a spin-orbit mediated non-radiative transition from the $uve$ spin-doublet to the *ue*² first excited state (purple dashed arrows in Fig. 1b). This mechanism is expected to manifest itself as a laser power-dependent $\gamma'_{3,4}$ rates during a continuous resonant or off-resonant optical excitation of the $V_{Si}^-$ defect further evidenced by our experimental observations.

**Experimental determination of the spin-dependent excited-state lifetimes**

Firstly, we measure the spin-dependent excited-state lifetimes using the experimental sequence depicted in Fig. 2a (see also Methods and ref. 21). A "repump" laser pulse is used to ensure that the $V_{Si}^-$ center is in the desired negative charge state. This is followed by a sub-lifetime short (1.5 ns) laser pulse at 916.5 nm, resonant with either the $O_1$ or $O_2$ transition. The fluorescence decay signal is recorded and subsequently fitted using a single exponential function as shown in Fig. 2b. We determine the excited-state lifetimes of bulk V2 centers as 6.1 ns and 11.3 ns for the $O_1$ and $O_2$ transitions, respectively.

Here, we note that the $O_2$ lifetime is almost twice as long as $O_1$, which is in stark contrast with the two nearly identical lifetimes of V1 center in 4H-SiC[21]. This indicates the V2 center has a much slower intersystem crossing for the $\pm 3/2$ spin sublevels associated with the $O_2$ transition (compared to the spin $\pm 1/2$ sublevels belonging to the $O_1$ optical transition). This also implies a significantly higher quantum efficiency for the $O_2$ transition which has been experimentally observed in a recent work with the V2 centers integrated into nanophotonic resonators[16].

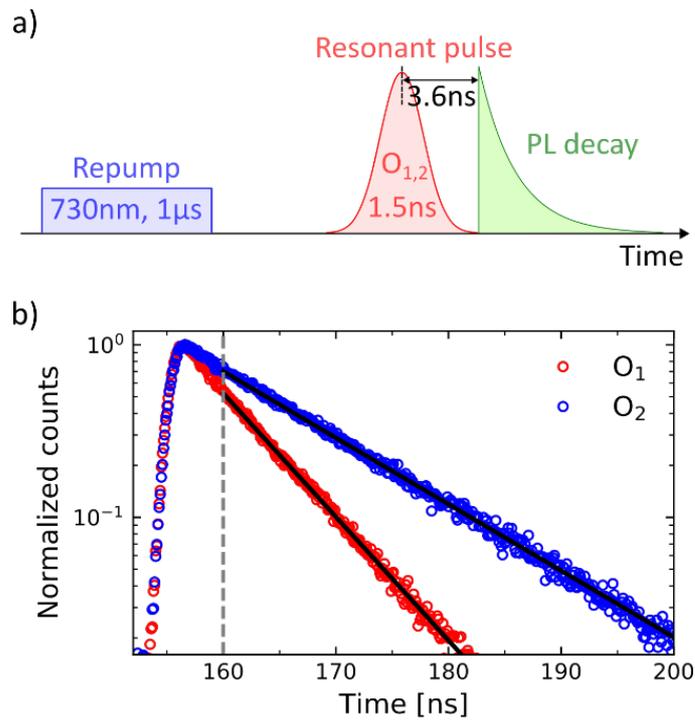

**FIG. 2. Spin-dependent excited-state lifetimes of a single V2 center.**

**a)** Measurement sequence consisting of charge-state initialization with off-resonant laser followed by $O_1$ or $O_2$ sub-lifetime short pulses, and fluorescence decay detection.

**b)** Fluorescence decay signals from the excited states. Solid black lines are single exponential fits.

**Probing the metastable-state dynamics of the V2 center**

We now investigate the predominant decay processes out of metastable states. This requires pumping of the system into the MSs followed by the determination of subsequent spin populations $p_{1/2}$ and $p_{3/2}$ within the spin subspaces of $m_s = \pm 1/2$ and $m_s = \pm 3/2$, respectively. For these studies, we take advantage of the following three key features:

First, it is known that a prolonged off-resonant excitation of $V_{Si}^-$ center pumps the system into the MS which is followed by a partial, non-complete spin polarization. It is assumed that the $m_s = \pm 1/2$ subspace shows a slightly higher population compared to the $m_s = \pm 3/2$ subspace[23].

Second, a prolonged resonant excitation along the $O_2$ or $O_1$ transitions can be used to achieve a (nearly) complete polarization into the $m_s = \pm 1/2$ or $m_s = \pm 3/2$ subspaces, respectively[10,21].

Third, the resonant excitation permits us to selectively read out the spin populations $p_{1/2}$ and $p_{3/2}$ in the subspaces $m_s = \pm 1/2$ and $m_s = \pm 3/2$.

In general, the non-equal quantum efficiencies of the $O_2$ or $O_1$ lead to different fluorescence signal strengths which makes the normalization of the population data challenging. For this reason, we use the fringe contrast of spin-Rabi oscillations providing a reliable way for the normalization. The related experimental sequence is shown in Fig. 3a.

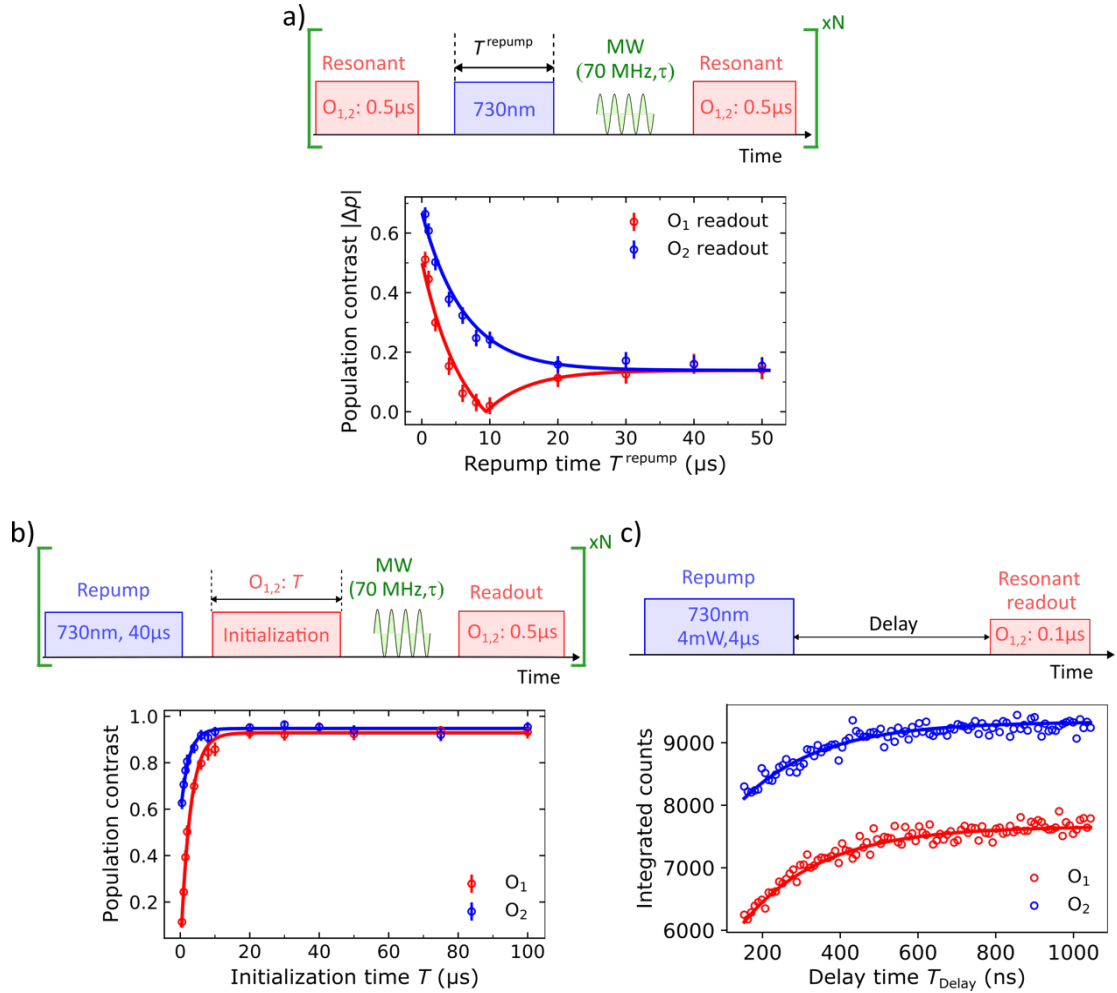

**FIG. 3. Decay dynamics of the metastable states probed by the spin-Rabi population contrast and delayed pulse measurements.**

**a)** Sequence for spin-Rabi contrast measurement for each point (i.e., each repump time $T^{\mathrm{repump}}$) and the evolution of the ground-state population contrast (absolute value) under resonant $O_1$ and $O_2$ readout lasers with increasing repump time.

**b)** The optical pumping fidelity with resonant laser defined by the population contrast with charge state initialized by the repump laser.

**c)** Sequence for delayed pulse measurement with varying delay times before resonant readout (upper). Photoluminescence counts integrated over the first 100 ns during resonant readout pulses as function of the delay time (lower). Solid lines are from simulation using all inferred rates summarized in Tab. I.

The sequence starts by predominantly initializing the system into the $m_s = \pm 1/2$ ($m_s = \pm 3/2$) spin subspace using ~0.5 µs long resonant excitation along the $O_2$ ($O_1$) transition. Subsequently, an off-resonant laser pulse, with varying duration of $T^{\text{repump}}$ and fixed power of 30 µW, is applied to increase the population in the MS, after which we allow the system to relax back to the ground states. To determine the *absolute* spin populations $p_{1/2}$ and $p_{3/2}$, we then perform spin-Rabi oscillations using a microwave drive at a frequency of 70 MHz, corresponding to the ground state splitting between the $m_s = \pm 1/2$ and $m_s = \pm 3/2$ subspaces. Finally, the spin population is read out by integrating the fluorescence intensity during a 0.5 µs short resonant laser pulse along the $O_2$ ($O_1$) transition. We note that the short readout pulse duration ensures that we exclude any complex dynamics stemming from ISCs, so that the fluorescence intensity signal is proportional to the ground state population. Overall, this sequence allows us to determine the ground state population contrast $\Delta p = (|p_{1/2} - p_{3/2}|)/(p_{1/2} + p_{3/2})$ as a function of the off-resonant laser pulse duration $T^{\text{repump}}$ (for more details, see Supplementary Note 1). The bottom inset of Fig. 3a shows the obtained experimental data. For short times of $T^{\text{repump}}$, the system has not yet reached an equilibrium (partial incomplete spin polarization into the $m_s = \pm 1/2$ subspace[23]). This is witnessed by a strong population contrast in both cases after initialization into the subspaces of either $m_s = \pm 1/2$ (blue) or $m_s = \pm 3/2$ (red). For $T^{\text{repump}} > 40$ µs, the population contrast reaches a steady state value of $\Delta p \sim 0.14$. Crucially, the behavior of the population contrast for both initializations is different. After initialization into $m_s = \pm 1/2$, we find a monotonic decay of $\Delta p$. In contrast for initialization into $m_s = \pm 3/2$, we observe a decay to $\Delta p \sim 0$ at $T^{\text{repump}} \sim 10$ µs followed by an increase to the steady state value $\Delta p \sim 0.14$. This is explained by the initial population $p_{3/2} \sim 1$ dropping to $p_{3/2} \sim 0.5$ after $T^{\text{repump}} \sim 10$ µs, and further decreasing to $p_{3/2} \sim 0.43$ for $T^{\text{repump}} > 40$ µs. In other words, our experimental results unambiguously determine that the $m_s = \pm 1/2$ subspace is preferably populated from the MSs with a ratio of 0.57/0.43 at an off-resonant laser power of 30 µW. It is important to mention that optical re-excitation/deshelving processes within the MSs are allowed by the selection rules (see Fig. 1b) and have been experimentally observed[34,35]. Re-excitation can drastically alter the spin population dynamics and as we show in the Supplementary Note1 we can achieve a ground state population ratio of 0.65/0.35 for an off-resonant laser power of 4 mW.

Having already established that the system reaches a steady state at 30 µW off-resonant pump power after $T^{\text{repump}} \sim 40$ µs, we now proceed to the high-fidelity spin initialization using a resonant excitation along the $O_1$ ($O_2$) transition. The related experimental sequence is shown in Fig. 3b (upper panel). An off-resonant laser pulse of 40 µs duration and 30 µW power establishes a steady state (0.57/0.43 ground state populations). Thereafter, a resonant laser pulse of duration $T$ and power 6 nW initiates the high-fidelity spin pumping. The absolute ground state spin populations are read out as before (spin Rabi oscillations, followed by a 0.1 µs spin-selective resonant laser excitation). The experimental results are shown in Fig. 3b (lower inset). For $T > 20$ µs, we reach high initialization fidelities of 95(1)% and 93(1)% into the spin subspaces $m_s = \pm 1/2$ (blue) or $m_s = \pm 3/2$ (red) under $O_2$ and $O_1$ optical pumping, respectively.

To further investigate the decay rates from the metastable (MS) states to the ground states (see Fig. 1b), we develop a delayed measurement scheme, as depicted in Fig. 3c. We use a 4 µs long off-resonant laser excitation pulse at 4 mW power, to trap most of the spin population in the MSs at the end of the pulse. This is due to the considerably longer lifetimes of the MSs compared to the excited-state lifetimes and the optical excitation rates. We then capture the MS depopulation dynamics by measuring the time-dependent increase of the ground state populations. To this end, we permit the system to relax to the ground state for a duration of $T_\text{Delay}$ before starting to integrate the fluorescence emission for 0.1 μs during a resonant excitation along the $O_1$ and $O_2$ transitions. We note that these measurements capture all the rates in and out of the MS states[24] which are populated by the incoming ISC rates ($\gamma_{1,2}, \gamma'_{1,2}$) and depopulated by the outgoing ones ($\gamma_{3,4}, \gamma'_{3,4}$). The related experimental data is shown in Fig. 3c (bottom inset). The graph also includes the simulated curves using the ISC model and rates given in Tab. I showing an excellent agreement between the experiment and theory. We note that these rates have been inferred through the MS dynamics probed until here and subsequent power-dependent resonant-excitation measurements which are discussed in the next section.

**Power-dependent resonant-excitation spin initialization measurements**

To obtain a comprehensive understanding of the spin initialization process through resonant excitation, we use the resonant laser power as an additional probing parameter ranging from 6 nW to 20 nW. We limit the maximum power to 20 nW to avoid excessive photo-ionization[26], as well as power broadening, which would result with the loss of spin selectivity[27]. As depicted in Fig. 4a, the measurement sequence consists of an off-resonant excitation pulse (30 μW, 40 μs) to initialize the $V_{Si}^-$ center into the negatively charged state and to initialize the ground state into slightly unbalanced spin populations (see Fig. 3a). Then, we selectively depopulate the spin subspaces $m_s = \pm 1/2$ or $m_s = \pm 3/2$ through continued resonant excitation along the $O_1$ (or $O_2$) optical transition while recording the fluorescence intensity for 40 μs. At all power levels, 40 μs resonant excitation is sufficiently long to completely depopulate the respective spin sublevels, as witnessed by the fluorescence signals reaching the noise level of the single-photon detectors. As shown in Fig. 4b, the time-dependent fluorescence intensity shows a tail that extends over to several microseconds. This indicates the involvement of a long-lived metastable state in the ISC, which is later corroborated by our rate results (see Tab. I).

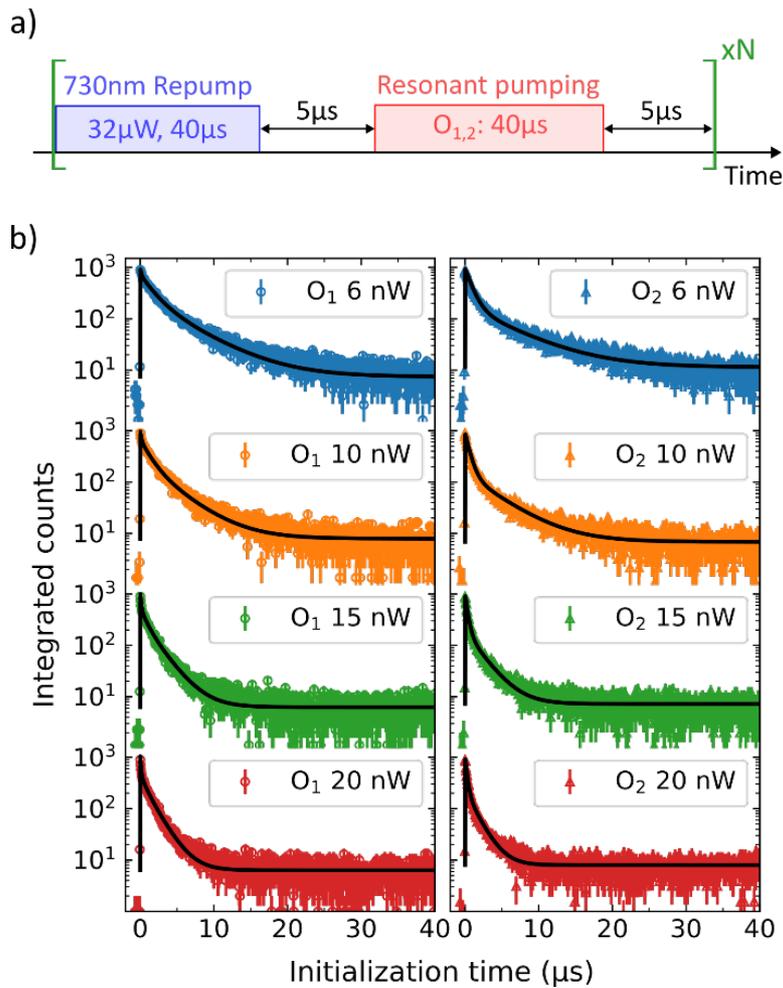

**FIG. 4. Time-dependent fluorescence decay under resonant excitation.**

**a)** Experimental sequence based on state initialization with an off-resonant repump laser, followed by spin-selective resonant pumping.

**b)** Measured fluorescence intensity signals for depopulation of both spin subspaces with the $O_1$ (left) and $O_2$ (right) lasers. The decay curves are measured and analyzed for varied resonant laser powers.

We now build a parameter optimization and fine-tuning algorithm that can be carried out over the density matrix master equation of the spin selective ISC model shown in Fig. 1b., constrained only by the measured excited state lifetimes (see Methods). The time dependent fluorescence decay data at all four powers and the delayed pulse measurement data are all simultaneously fitted using this algorithm. The resulting fit curves for the fluorescence decays are presented in Fig. 4. showing excellent agreement with the experimental data. Individual analysis of the resonant initialization and delayed pulse measurement leads to a differing number of metastable states. Our ISC model presented in Fig. 1b leads to an excellent agreement with both measurements in Fig. 3 and 4 and involves minimum number of metastable states for accurately describing the experimental data. The resulting transition rates are summarized in Tab. I. In accordance with the model in Fig. 1b, we find two effective metastable states playing a significant role in the ISC. The first metastable state (MS$_1$) consists of the $ve^2$ and $e^3$ spin-doublet states in which the fast dynamic relaxation ($\gamma_{n1} \gg \gamma_i, \gamma_i'$) of $ve^2$ onto $e^3$ is captured within. The second effective metastable state (MS$_2$) is formed from the $v^2e$ spin-doublet state. Due to the deshelving of $v^2e$ spin doublet to $uve$ spin doublet under optical excitation (see Fig. 1b), the MS$_2$ lifetime shows a significant power dependence during resonant excitation measurements. Using the deshelving model and taking 20 nW as a reference power, from each of the remaining fit curves we theoretically infer the rest of the resonant excitation powers. The calculated powers are well within the experimental error tolerances showing excellent agreement (see Tab. I) across all powers and providing further evidence of the deshelving mechanism.

At this point we highlight key differences between the V1 and V2 centers in 4H-SiC. In our previous work, we showed that V1 center has a metastable state with ∼ 200 ns lifetime[21]. This is very similar to the MS$_1$ lifetime of the V2 center, as we showed in this work. However, the V2 center shows an additional long-lived MS$_2$ metastable state (up to ~ 3 μs at low excitation powers), for which no such evidence was observed for V1. We explain this difference by the PJT effect which results in a strong vibronic mixing between the MS$_1$ ($ve^2$) and MS$_2$ ($v^2e$) states. For the V1 center with near-$T_d$ symmetry, this effectively results in a single metastable state due to the increased degeneracy between the $v$ and $e$ MOs. On the other hand, for V2 centers, such degeneracy is removed by the much more distorted local symmetry ($C_{3v}$) along the c-axis[28], suppressing any mixing. The weaker PJT effect for V2 centers is additionally confirmed by our recent work, which showed that V2 centers maintain narrow optical linewidths at significantly higher temperatures compared to V1 centers[28]. The very long lifetime of MS$_2$ also affects the behavior of V2 color centers under off-resonant excitation. Especially at high laser powers, Stokes excitation can lead to another depletion channel for MS$_2$ which eventually reduces the effective lifetime of the entire metastable state manifold to the lifetime of the MS$_1$. This behavior is experimentally corroborated by our high-repump-power results in Fig. 3c, as well as previous room-temperature investigations[34,35].

**TABLE I. Radiative and non-radiative transition lifetimes of V2 color centers in 4H-SiC.**

**(Top)** The MS$_1$ ($e^3$) metastable state lifetime is inferred according to the ISC mechanism described in Fig. 1b and Fig. 1c-iv.

**(Bottom)** The MS$_2$ ($v^2e$) metastable state lifetime is found to be dependent on the optical excitation power due to the *uve* spin-doublet shelving state shown in Fig. 1b. The experimental resonant excitation powers and the theoretically inferred ones (20 nW as a reference power for calculation) are shown.

| Transitions | $1/\gamma_r$ | $1/\gamma_1$ | $1/\gamma_1'$ | $1/\gamma_2$ | $1/\gamma_2'$ | $1/\gamma_3$ | $1/\gamma_4$ |
|---|---|---|---|---|---|---|---|
| Lifetimes (ns) | 17.84 | 11.05 | 56.75 | 130.59 | 41.02 | 250.72 | 1035.35 |
| MS$_1$ Lifetime ($1/\gamma_3 + 1/\gamma_4$) | | | | | | 201.84 ns | |

| Resonant excitation power (nW) | Theoretically inferred power (nW) | $1/\gamma_3'$ (=$1/\gamma_4'$) (ns) | MS$_2$ Lifetime (ns) |
|---|---|---|---|
| 6 | 5.58 | 5928.73 | 2964.37 |
| 10 | 8.44 | 4377.85 | 2188.93 |
| 15 | 15.01 | 2170.80 | 1085.40 |
| 20 | Reference | 1481.69 | 740.85 |

**Discussion**

From our measured rates for V2 centers in 4H-SiC, the higher quantum efficiency of the $O_2$ transition with spin $|m_s| = 1/2$ leads to a higher cooperativity (see Supplementary Note 2) when integrated in nanophotonic resonators, which has been recently observed[16]. Here, we develop protocols for the generation of time-bin entangled multi-photon states from a single V2 center that can take advantage of this high quantum efficiency of the $O_2$ transition. We perform a detailed analysis of our protocols and consider multiple sources of imperfections including both spin conserving (i.e., excited-state phonon scattering, imperfect excitation) and spin-flip (due to ISC mechanism) errors.

Two ground-state spins $|g_1\rangle = |-1/2\rangle$ and $|g_2\rangle = |-3/2\rangle$ are chosen as entanglers for the Greenberger-Horne-Zeilinger (GHZ) and one-dimensional cluster states. Our protocol is adapted from a similar concept initially developed for quantum dots[29] and consists of periodic optical drive of $O_2$ and coherent microwave control of ground-state spins. The Kramer's degenerate ground-state spins are further split into four sublevels (see Fig. 5a) by applying a sufficiently large magnetic field (e.g., $B = 5$ mT) for longer spin coherence time[30]. The entire protocol shown in Fig. 5b involves the following steps:

i. The ground state spin is initialized to $|g_1, 0\rangle + |g_2, 0\rangle$ (normalization omitted w.l.o.g.) by resonant optical pumping of $O_1$ transition during a continuous microwave driving ($I_{MW}$) of the spin $|+3/2\rangle \leftrightarrow |+1/2\rangle$ transition followed by a microwave $\pi/2$ pulse ($S_{MW}^{\pi/2}$) resonant with $|-3/2\rangle \leftrightarrow |-1/2\rangle$.

ii. An optical $\pi$ pulse resonantly excites the $O_2$ transition resulting with the spontaneous emission of the first ZPL photon into an early time bin: $|g_1, 0\rangle + |g_2, 1_E\rangle$.

iii. The spin states are swapped via $S_{MW}^{\pi}$ pulse: $|g_2, 0\rangle + |g_1, 1_E\rangle$.

iv. Upon resonant excitation of $O_2$ with a second $\pi$ pulse, another photon is emitted into a late time bin: $|g_2, 1_L\rangle + |g_1, 1_E\rangle$.

v. For a GHZ state generation, a final $R = S_{MW}^{\pi/2}$ pulse (i.e., X-gate) is applied resulting with $|g_1, 1_L\rangle + |g_2, 1_E\rangle$. Similarly, for a cluster state generation, the Pauli-X gate can be replaced by a Hadamard gate ($R = H = XY^{1/2}$) resulting with a generator $C^\dagger = |g_+\rangle\langle g_2|a_E^\dagger + |g_-\rangle\langle g_1|a_L^\dagger$ for each period with $|g_\pm\rangle = (|g_1\rangle \pm |g_2\rangle)/\sqrt{2}$. By repeating one period (from ii to v) of this protocol $N$ times we obtain an $N$-photon GHZ state or a cluster state depending on the final gate operation.

The dephasing of the excited states for V2 centers induced by acoustic phonons was shown to be negligible by the preservation of narrow PLE linewidths up to 20 K[28]. Therefore, the fidelity related to phonon-induced pure dephasing, defined[29] as $F_p^{GHZ,C} = 1 - N\gamma_d/(\gamma_r + 2\gamma_d)$, is close to 1 with $\gamma_d \ll \gamma_r$. We also consider excitation errors in the resonant driving pulses that induce undesired stimulated photon emissions (weak, long pulse) and detuned $O_1$ transition (strong, short pulse) given[29] as $F_{ex} = 1 - N\left(\frac{\sqrt{3}\pi}{8}\right)\gamma_r/\Delta$. With our measured radiative decay rate and ~1 GHz separation between $O_1$ and $O_2$ transitions, the pulse timing for a square $\pi$-pulse optimized upon $2\pi$-rotation of the detuned transition is found to be 0.9 ns for a 3-photon GHZ/cluster state resulting in an excitation fidelity of 74.3%. The fidelity of the GHZ and cluster states with branching errors (emission into phonon sideband and ISC) are the same for the protocol shown in Fig.5 and given as $F_{br}^{GHZ,C} = (P_{O_2})^N$ conditioned upon a successful detection without any post-processing. $P_{O_2}$ is the ZPL emission probability for the $O_2$ radiative transition and defined as $(P\alpha\gamma_r)/((1-\alpha)\gamma_r + P\alpha\gamma_r + \gamma_2 + \gamma_2')$ in terms of Debye-Waller factor $\alpha$~9% [28], Purcell factor $P$, and the rates inferred in Tab. 1. We find $F_{br}^{GHZ,C} = 0.06$ without any cavity Purcell enhancement with the above optimized pulse timing. The cavity enhancement of radiative lifetime improves $F_p^{\text{GHZ,C}}$ and $F_{\text{br}}^{\text{GHZ,C}}$, but it degrades $F_{\text{ex}}$ slightly as the excited state radiative lifetime is modified as $\gamma_r \to (1-\alpha)\gamma_r + P\alpha\gamma_r$. To maximize the overall fidelity defined as $F_t = F_p F_{\text{ex}} F_{\text{br}}$, we calculate the optimized Purcell factors for each state size as shown in Fig. 5c. The primary limitation to the total fidelities shown here comes from the competition between the excitation and branching ratio errors as $F_{\text{br}}$ increases while $F_{\text{ex}}$ decreases with larger Purcell factors. To reach a total fidelity of 50%, GHZ/cluster states of maximum 3 photons are feasible. We calculate the minimum required Purcell factors to reach $F_t = 50\%$ (Fig. 5d) for various state sizes compatible with existing nanophotonic resonators[15,16,31]. Photon states of larger size (> 3 photons) can be achieved by increasing the excited-state ZFS (∼Δ) via applied strain allowing for much larger Purcell enhancements to be applied[32].

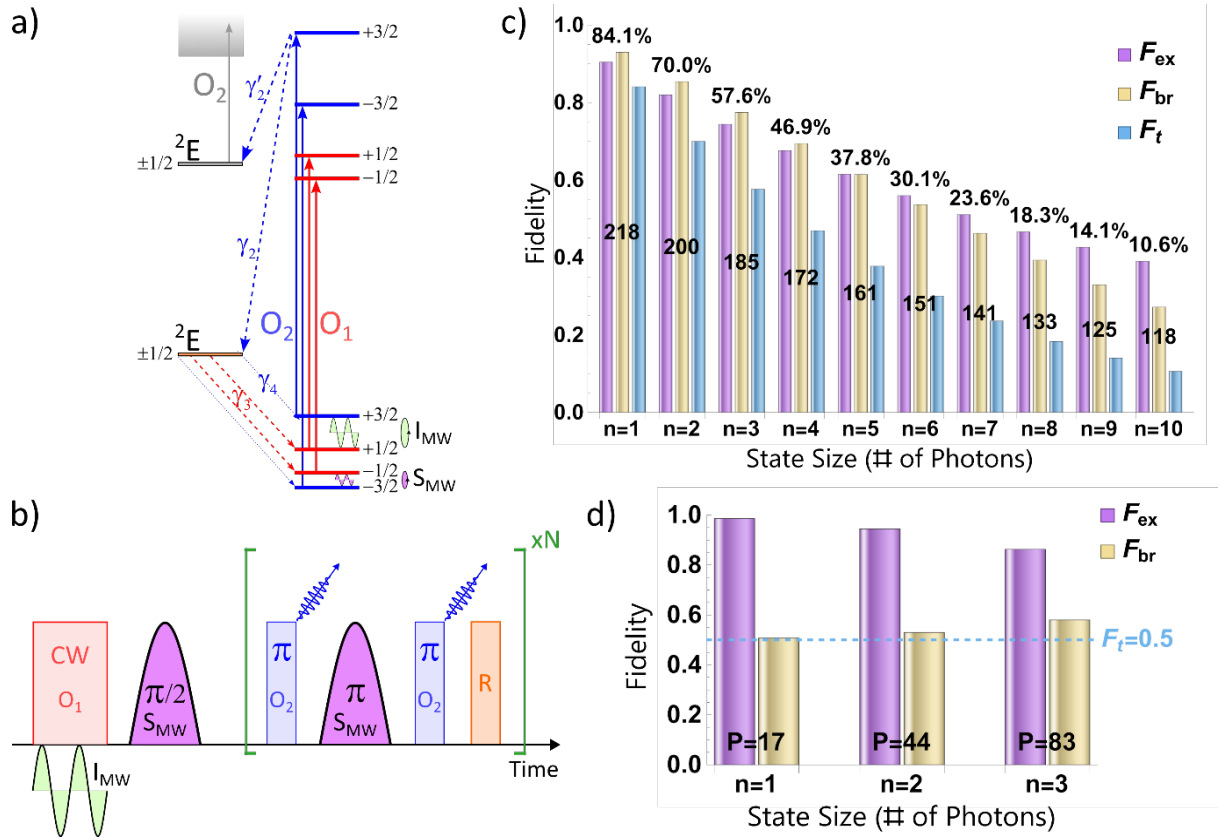

**FIG. 5. Generation of entangled multi-photon GHZ and cluster states.**

**a)** Relevant energy scheme of $V_{Si}^-$ used in the protocol.

**b)** Pulse sequences described in the main text.

**c)** Total estimated fidelity of GHZ and cluster states with sizes up to n=10 photons as a product of excitation and branching ratio errors. Optimized Purcell enhancement factors for each state size are shown on the bars.

**d)** Minimum Purcell factor requirements to reach a total fidelity of 0.5 for GHZ and cluster states up to 3 photons.

**Summary**

In summary, we have established the complete electronic fine structure and intrinsic spin dynamics of the V2 silicon vacancy centers in 4H-SiC unraveling all the previously unknown spin-dependent radiative and non-radiative decay rates, ISC, and deshelving mechanisms. The mechanisms identified here successfully explain several previous measurements done with V2 centers including anti-Stokes excitation[33], ODMR with off-resonant readout[35], as well as autocorrelation of resonator-integrated defects[16]. Our work also explains the main differences between V1 and V2 centers in 4H-SiC. The newly found understanding of the complete spin-optical dynamics of the V2 center provides the critical engineering guidelines towards its integration into nanophotonic enhancement structures, such as waveguides and resonators. To showcase this, we additionally proposed realistic protocols for generating time-bin entangled multi-photon GHZ and cluster states, taking advantage of the high quantum efficiency of the $O_2$ transition. We provided in-depth analysis of state fidelities, optimization of pulse timings, and minimum Purcell enhancement requirements for generating GHZ/cluster states of various sizes. We show that 2-photon GHZ and cluster states can be readily realized with existing SiC nanophotonic resonators, whereas higher photon-number states would require further improvements. In this sense, we believe that phonon/strain engineering of V2 centers will become necessary to suppress excitation errors by increasing the exited state zero-field splitting, and to improve the overall branching ratio.

Overall, our studies provide a holistic summary on the intrinsic spin-optical dynamics of the V2 center in 4H-SiC. This now permits defining ideal experimental protocols and routines for maximizing the performance in various quantum technology applications, as well as optimizing the optical performance of V2 centers via integration into nanophotonic resonators. Additionally, our methods can be straightforwardly adapted to improve the understanding of the internal dynamics of other color centers.

## METHODS

### Experimental setup

All the experiments are performed with a home-built scanning confocal microscope at 5.5 K in a closed-cycle cryostat (Montana Instruments). The resonant excitation of the single V2 color center uses a tunable single-frequency diode laser (Toptica DLC DL pro) at 916.5 nm. The acousto-optic modulator (Gooch&Housego) and electro-optic amplitude modulator (Jenoptik) enable continuous-wave and pulsed excitation resonantly. A custom-made diode laser is employed for off-resonant excitation at 730 nm. A polarization-maintaining fiber combines all excitation lasers which are focused onto the sample by a microscope objective (Zeiss EC Epiplan-Neofluar x 100, NA=0.9). The scanning of the sample is enabled by a fast-steering mirror (Mad City Labs). The fluorescence emission is collected at phonon-side bands (940 nm-1033 nm) by a superconducting nanowire single photon detector (Photon Spot). The fabrication of the 4H-SiC sample, defect generation, and development of a solid immersion lens have been described in our previous work[10].

### Density matrix master equation parameter optimization

The resonant PL decay of the V2 defect can be accurately modeled by using the fine structure and ISC model in Fig. 1b. The spin-selective fluorescence signal corresponds to the time-dependent excited state populations, that are calculated using the following master equation, $d\rho/dt = -\frac{i}{\hbar}[H_0, \rho] + \gamma_r \sum_{i=1}^{2} L(A_r^i) + \sum_{i=1}^{4} \gamma_i L(A_{ms_1}^i) + \sum_{i=1}^{4} \gamma_i' L(A_{ms_2}^i)$. The radiative and non-radiative decay processes are represented by the Lindblad super-operators, $L(O)$. The $H_0$ is the Hamiltonian constructed from optically driven spin $\pm 1/2$ and $\pm 3/2$ ground and excited states, $H_0 = (D_g - D_e - \delta_L)(|gs_{1/2}\rangle\langle gs_{1/2}| - |es_{1/2}\rangle\langle es_{1/2}|) - (D_g - D_e - \delta_L)(|gs_{3/2}\rangle\langle gs_{3/2}| - |es_{3/2}\rangle\langle es_{3/2}|) + [\Omega_L(|gs_{1/2}\rangle\langle es_{1/2}| +$

$|gs_{3/2}\rangle\langle es_{3/2}|) + h.c.]$ in the rotating frame of the laser with power dependent Rabi frequency $\Omega_L$ and detuning $\delta_L = \omega_L - \omega_{ZPL}$. The radiative decays, $A_r^i = |gs_i\rangle\langle es_i|$, are governed by the same radiative decay rates $\gamma_r$ for both O$_1$ and O$_2$ transitions. The non-radiative ISC decays in and out of each metastable state (ms$_1 = ve^2$ and ms$_2 = v^2e$) are given by $A_{ms_1}^i = |ms_1\rangle\langle \varphi_i|$ and $A_{ms_2}^i = |ms_2\rangle\langle \varphi_i|$ in which $\varphi_{\{i=1,2,3,4\}} = \{es_{1/2}, es_{3/2}, gs_{1/2}, gs_{3/2}\}$ after the effective rate simplification with $\gamma_{n1} \gg \gamma_{3,4}$ and $\gamma_{n2} \ll \gamma'_{3,4}$. We use a custom-built parameter optimization algorithm based on both Nelder-Mead and differential evolution numerical nonlinear optimization methods for simultaneously fitting the excited state population solutions of separate master equations at four different laser powers with the time dependent PL decay measurement data. At each trial, a secondary simplified master equation reflecting the 100ns integration window for pulse sequence in Fig. 1c is used to also evaluate the fit quality of each rate solution with the delayed pulse measurement data.

**References**

[1] Awschalom, D. D., Hanson, R., Wrachtrup, J. & Zhou, B. B. Quantum technologies with optically interfaced solid-state spins. *Nat. Photonics* **12**, 516–527 (2018).

[2] Bradley, C. E. *et al.* A ten-qubit solid-state spin register with quantum memory up to one minute. *Phys. Rev. X* **9**, 031045 (2019).

[3] Pompili, M. *et al.* Realization of a multinode quantum network of remote solid-state qubits. *Science* **372**, 259–264 (2021).

[4] Stas, P.-J. *et al.* Robust multi-qubit quantum network node with integrated error detection. *Science* **378**, 557–560 (2022).

[5] Castelletto, S. & Boretti, A. Silicon carbide color centers for quantum applications. *J. Phys. Photonics* **2**, 022001 (2020).

[6] Röder, R. *et al.* Transition metal and rare earth element doped zinc oxide nanowires for optoelectronics. *Phys. Status Solidi (b)* **256**, 1800604 (2019).

[7] Vaidya, S., Gao, X., Dikshit, S., Aharonovich, I. & Li, T. Quantum sensing and imaging with spin defects in hexagonal boron nitride. *Adv. Phys. X* **8**, 2206049 (2023).

[8] Widmann, M. *et al.* Coherent control of single spins in silicon carbide at room temperature. *Nat. Mater.* **14**, 164–168 (2015).

[9] Christle, D. J. *et al.* Isolated electron spins in silicon carbide with millisecond coherence times. *Nat. Mater.* **14**, 160–163 (2015).

[10] Nagy, R. *et al.* High-fidelity spin and optical control of single silicon-vacancy centres in silicon carbide. *Nat. Commun.* **10**, 1954 (2019).

[11] Morioka, N. *et al.* Spin-controlled generation of indistinguishable and distinguishable photons from silicon vacancy centres in silicon carbide. *Nat. Commun.* **11**, 2516 (2020).

[12] Bourassa, A. *et al.* Entanglement and control of single nuclear spins in isotopically engineered silicon carbide. *Nat. Mater.* **19**, 1319–1325 (2020).

[13] Anderson, C. P. *et al.* Five-second coherence of a single spin with single-shot readout in silicon carbide. *Sci. Adv.* **8**, eabm5912 (2022).

[14] Babin, C. *et al.* Fabrication and nanophotonic waveguide integration of silicon carbide colour centres with preserved spin-optical coherence. *Nat. Mater.* **21**, 67–73 (2022).

[15] Crook, A. L. *et al.* Purcell enhancement of a single silicon carbide color center with coherent spin control. *Nano Lett.* **20**, 3427–3434 (2020).

[16] Lukin, D. M. *et al.* Two-emitter multimode cavity quantum electrodynamics in thin-film silicon carbide photonics. *Phys. Rev. X* **13**, 011005 (2023).

[17] Day, A. M., Dietz, J. R., Sutula, M., Yeh, M. & Hu, E. L. Laser writing of spin defects in nanophotonic cavities. *Nat. Mater.* (2023).

[18] Lukin, D. M. *et al.* 4H-silicon-carbide-on-insulator for integrated quantum and nonlinear photonics. *Nat. Photon.* **14**, 330–334 (2020).

[19] Greenberger, D. M., Horne, M. A. & Zeilinger, A. *Going Beyond Bell's Theorem*, 69–72 (Springer Netherlands, Dordrecht, 1989).

[20] Schwartz, I. *et al*. Deterministic generation of a cluster state of entangled photons. *Science* **354**, 434–437 (2016).

[21] Morioka, N. *et al.* Spin-optical dynamics and quantum efficiency of a single V1 center in silicon carbide. *Phys. Rev. Appl.* **17**, 054005 (2022).

[22] Soykal, O. O., Dev, P. & Economou, S. E. Silicon vacancy center in 4H-SiC: Electronic structure and spin-photon interfaces. *Phys. Rev. B* **93**, 081207 (2016).

[23] Simin, D. *et al.* All-optical dc nanotesla magnetometry using silicon vacancy fine structure in isotopically purified silicon carbide. *Phys. Rev. X* **6**, 031014 (2016).

[24] Robledo, L. *et al.* Spin dynamics in the optical cycle of single nitrogen-vacancy centres in diamond. *New J. Phys.* **13**, 025013 (2011).

[25] Widmann, M. *et al.* Electrical charge state manipulation of single silicon vacancies in a silicon carbide quantum optoelectronic device. *Nano Lett.* **19**, 7173–7180 (2019).

[26] Niethammer, M. *et al.* Coherent electrical readout of defect spins in silicon carbide by photoionization at ambient conditions. *Nat. Commun.* **10**, 5569 (2019).

[27] Banks, H. B. *et al.* Resonant optical spin initialization and readout of single silicon vacancies in 4H-SiC. *Phys. Rev. Appl.* **11**, 024013 (2019).

[28] Udvarhelyi, P. *et al.* Vibronic states and their effect on the temperature and strain dependence of silicon-vacancy qubits in 4H-SiC. *Phys. Rev. Appl.* **13**, 054017 (2020).

[29] Tiurev, K. *et al*. Fidelity of time-bin-entangled multiphoton states from a quantum emitter. *Phys. Rev. A* **104**, 052604 (2021).

[30] Childress, L. et al. Coherent dynamics of coupled electron and nuclear spin qubits in diamond. *Science* **314**, 281–285 (2006).

[31] Bracher, D. O., Zhang, X. & Hu, E. L. Selective Purcell enhancement of two closely linked zero-phonon transitions of a silicon carbide color center. *Proc. Natl. Acad. Sci. U.S.A.* **114**, 4060–4065 (2017).

[32] Falk, A. L. et al. Electrically and mechanically tunable electron spins in silicon carbide color centers. *Phys. Rev. Lett.* **112**, 187601 (2014).

[33] Wang, J.-F. et al. Robust coherent control of solid-state spin qubits using anti-Stokes excitation. *Nat. Commun*. **12**, 3223 (2021).

[34] Fuchs, F. et al. Engineering near-infrared single-photon emitters with optically active spins in ultrapure silicon carbide. *Nat. Commun*. 6, 7578 (2014).

[35] Singh, H. et al. Characterization of single shallow silicon-vacancy centers in 4H-SiC. *Phys. Rev. B* **107**, 134117 (2023).

**Contributions**

D.L., F.K., Ö.O.S and J.W. conceived and designed the experiments. D.L. and F.K. performed the experiments. D.L. and Ö.O.S. analyzed the data. Ö.O.S. developed the theoretical modeling and simulations. D.L. and Ö.O.S. carried out the numerical simulations. J.U.-H. provided the SiC sample. T.O. and N.T.S. irradiated the SiC with electrons for defect generation. F.K., V.B., E.H., T.S. and J.W. assisted the data analysis. D.L., F.K. and Ö.O.S. wrote the manuscript with helpful inputs from all the authors.

**Acknowledgements**

We acknowledge fruitful discussions with Petr Siyushev, Jianpei Geng, Naoya Morioka, Daniil Lukin and Melissa Guidry. D.L., F.K., V.B., E.H., T.S. and J.W. acknowledge support from the European Commission for the Quantum Technology Flagship project QIA (Grant agreements 101080128 and 101102140), the German ministry of education and research for the projects InQuRe, QR.X, and Spinning (BMBF, Grants No. 16KIS1639K, No. 16KISQ013, and No. 13N16219), as well as the Ministerium für Wirtschaft, Arbeit und Tourismus Baden-Württemberg for the project SPOC (Grant Agreement No. QT-6). F.K., J.U.H, and J.W. acknowledge support from the European Commission through the QuantERA project InQuRe (Grant Agreements No. 731473 and No. 101017733). J.U.H. acknowledges support from the Swedish Research Council under VR Grant No. 2020-05444. N.T.S. and J.U.H. acknowledge support from EU H2020 project QuanTELCO (Grant No. 862721), and the Knut and Alice Wallenberg Foundation (Grant No. KAW 2018.0071). Ö.O.S. acknowledges support from Booz Allen Hamilton Inc. and QuaSar (No. 23RDP001). T.O. acknowledges funding from the Japan Society for the Promotion of Science via the grant JSPS KAKENHI 21H04553, as well as the Japan Science and Technology Agency for funding within the MEXT Q-LEAP program via the grant JPMXS0118067395.

# Supplementary Information for:

## The silicon vacancy centers in SiC: determination of intrinsic spin dynamics for integrated quantum photonics

Di Liu[1], Florian Kaiser[2,#], Vladislav Bushmakin[1], Erik Hesselmeier[1], Timo Steidl[1],

Takeshi Ohshima[3,4], Nguyen Tien Son[5], Jawad Ul-Hassan[5], Öney O. Soykal[6,*] and Jörg Wrachtrup[1]

**Affiliations:**

[1] 3rd Institute of Physics, IQST, and Research Center ScoPE, University of Stuttgart, Germany

[2] Luxembourg Institute of Science and Technology, Luxembourg

[3] National Institutes for Quantum Science and Technology (QST), 1233 Watanuki, Takasaki, Gunma 370-1292, Japan

[4] Department of Materials Science, Tohoku University, 6-6-02 Aramaki-Aza, Aoba-ku, Sendai 980-8579, Japan

[5] Department of Physics, Chemistry and Biology, Linköping University, Sweden

[6] Booz Allen Hamilton Inc., USA (current affiliation: Photonic Inc., CA)

**Email correspondence:**

* Öney O. Soykal, osoykal@photonic.com
# Florian Kaiser, florian.kaiser@list.lu

**Supplementary Note 1: Extraction of ground-state population contrast**

Without applying an external static magnetic field along c-axis, the Kramer's degenerate ground states are split into spin-3/2 and -1/2 subspaces by the zero-field splitting (ZFS) of 70 MHz due to spin-spin interactions. To coherently drive the transition between the two spin subspaces, sufficient microwave (MW) power is sent into the system so that the MW $\pi$-pulse is about 100 ns long. Under this condition, the population oscillation between the ground states is probed by $O_1$ or $O_2$ resonant pulses during the MW drive. By fitting the oscillation fringe (also known as spin Rabi) with $I_{mean} \cdot \left[\frac{I_{max}-I_{min}}{I_{max}+I_{min}} \cdot \cos(\omega(t-t_0)) + 1\right]$ where $I_{mean} = (I_{max} + I_{min})/2$, the population contrast (defined in the main text) proportional to $\frac{I_{max}-I_{min}}{I_{max}+I_{min}}$ is extracted. This procedure is followed for each repump time. In Fig. S1 the Rabi oscillations are shown for the shortest and longest repump time applied in the experimental sequence. With increasing repump time, the change in the sign of the population contrast is also visible in the phase change of the Rabi oscillations comparing the two red dotted plots.

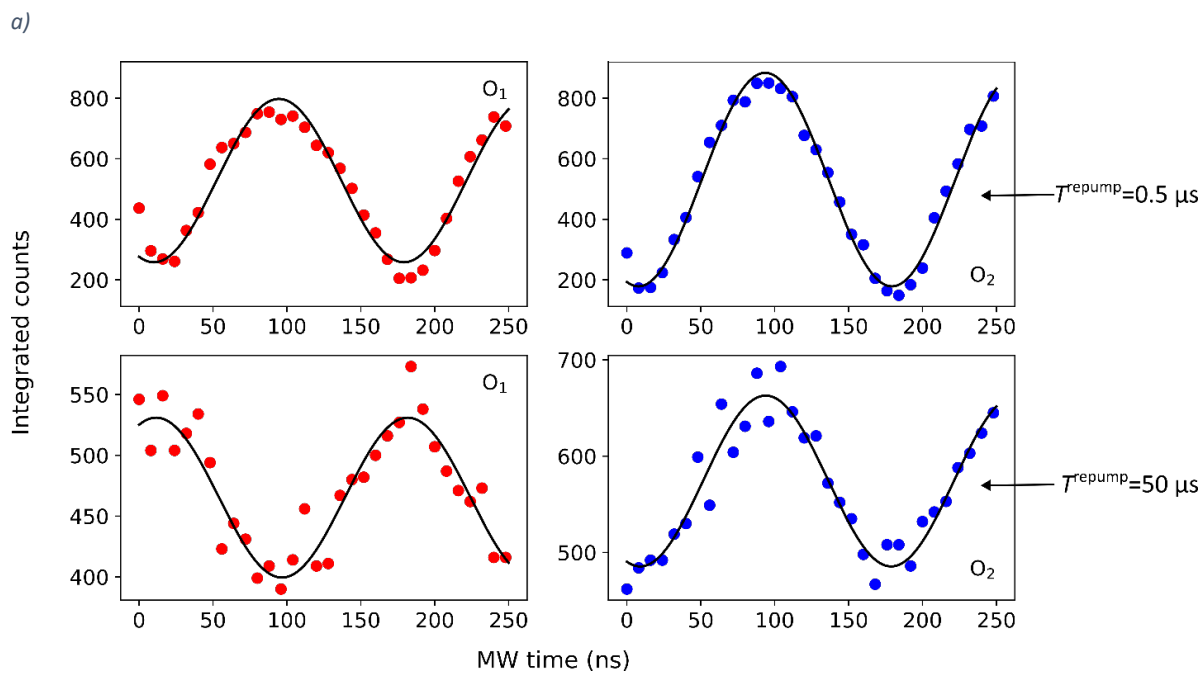

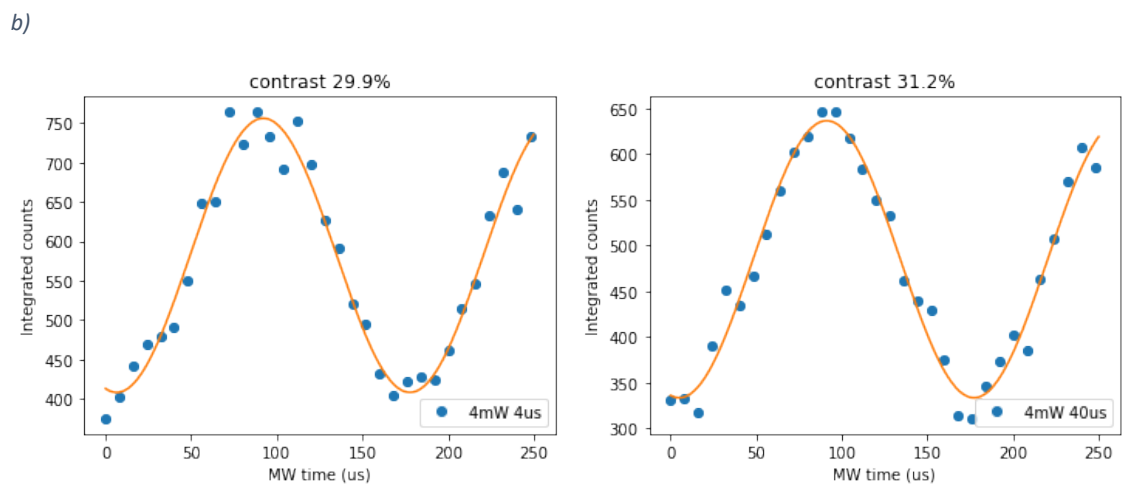

**Fig. S1.: Spin Rabi oscillations.**

**a)** Spin Rabi oscillations between the ground-state spin-3/2 and -1/2 subspaces split by the ZFS shown for the shortest and longest repump time applied in the population contrast measurement (see Fig. 5ab in the main text). Solid lines are fitting curves for extraction of the population contrast.

**b)** Spin Rabi oscillations measured as in a) with 4 mW repump power.

**Supplementary Note 2: Expression of coorperativity ($C$) using Purcell factor ($P$), quantum efficiency ($\eta$) and Debye-Waller factor ($\alpha$)**

The quantum efficiency of a quantum emitter is defined by the ratio between the radiative decay rate and the sum of all decay rates from the excited state, that is, the inverse of the excited-state lifetime: $\eta = k_\text{r} \cdot \tau_\text{ex}$. The Debye-Waller factor is given by the fraction of emission into the zero-phonon line (ZPL): DWF=$k_\text{ZPL}/k_\text{r} = (k_\text{ZPL} \cdot \tau_\text{ex})/\eta$. In nanophotonic cavity, the ZPL decay rate is enhanced by Purcell enhancement $k'_\text{ZPL} = p \cdot k_\text{ZPL}$. Using the definition of cooperativity which compares the enhanced ZPL decay rate and all the decay rates without cavity effect[1], the coorperativity is expressed in terms of the Purcell factor, quantum efficiency and the Debye-Walle factor of the bulk emitter as $C = p \cdot \eta \cdot \text{DWF}$.

**References:**

[1] Janitz, E., Bhaskar, M. K. & Childress, L. Cavity quantum electrodynamics with color centers in diamond. Optica **7**, 1232–1252 (2020).